\documentclass[final,5p,times,twocolumn,numbers,sort, compress]{elsarticle}
\usepackage{amssymb}
\usepackage{lipsum}
\usepackage{xcolor}
\usepackage{mathtools}
\usepackage{amsmath}
\usepackage[utf8]{inputenc}
\usepackage{subfig}
\usepackage{float}

\journal{Journal of Statistical Mechanics: Theory and Experiment}

\begin{document}

\begin{frontmatter}

\title{Scaling invariance for the diffusion coefficient in a dissipative standard mapping}

\author{$^1$Edson D. Leonel, $^1$C\'elia M. Kuwana, $^2$Diego F. M. Oliveira}
\address{{$^1$Departamento de F\'isica, UNESP - Universidade Estadual Paulista, Av. 24A, 1515, Bela Vista, Rio Claro, 13506-900, Sao Paulo, Brazil\\
$^2$School of Electrical Engineering and Computer Science - University of North Dakota, Grand Forks, Avenue Stop 8357, 58202, North Dakota, USA.}}

\begin{abstract}

The unbounded diffusion observed for the standard mapping in a regime of high nonlinearity is suppressed by dissipation due to the violation of Liouville's theorem. The diffusion coefficient becomes important for the description of scaling invariance particularly for the suppression of the unbounded action diffusion. When the dynamics start in the regime of low action, the diffusion coefficient remains constant for a long time, guaranteeing the diffusion for an ensemble of particles. Eventually, it evolves into a regime of decay, marking the suppression of particle action growth. We prove it is scaling invariant for the control parameters and the crossover time identifying the changeover from the constant domain, leading to diffusion, for a regime of decay marking the saturation of the diffusion, scales with the same critical exponent $z=-1$ for a transition from bounded to unbounded diffusion in a dissipative time dependent billiard system.
\end{abstract}


\end{frontmatter}

%\tableofcontents



\section{Introduction}
\label{introduction}

The movement of particles from a regime of high concentration to one of low concentration is called diffusion \cite{b1}. It is proven to be a crucial phenomenon appearing in numerous systems and areas in nature \cite{p1,p2,p3}, ranging from simple experiments such as the dispersal of ink in water \cite{b2}, the dissemination of pollutants in the air \cite{p4,p5,p6}, in oceans \cite{b3}, the dispersion of spacecraft and dead satellites orbiting the earth \cite{p7}, the spread of seeds in a forest \cite{p8,p9}, diffusion in medicine for medical treatment \cite{p10,p11,p12} or disease \cite{p13} such as the Covid \cite{p14,p15,p16} and Dengue \cite{p17,p18} and even Malaria \cite{p19} among many others standard processes observed in everyday life.

Even though it is a simple process, diffusion often has an intricate behavior, mainly when external forces or specific boundary conditions are considered \cite{b4}. Its background investigation comes from a phenomenological proposal by Fick in his first Fick's law \cite{b5}. It assumes that a flux of particles, producing then a current of particles, evolves from regions of high concentration to areas of low concentration. The magnitude of the flux is proportional to the concentration gradient, implying that the particles will move from a region of high concentration to an area of low concentration due to a concentration gradient. A quantity that makes a direct connection between these two assumptions is called the diffusion coefficient $D$. It therefore turns to be one of the main observables we will investigate and describe in this paper. 

We consider a dissipative standard model \cite{p20,p21,b6} to concentrate on discussing the diffusion coefficient. It is constructed as a two-dimensional nonlinear mapping $T$ for the variables action $I$ and angle $\theta$ relating them from the instant $n$ to $(n+1)$, i.e. $T(I_n,\theta_n)=(I_{n+1},\theta_{n+1})$. Depending on the control parameters of the mapping, it is known that unbounded diffusion of the action variable is observed \cite{b6} and that a dissipative term introduced to the equations is enough to suppress such a diffusion \cite{p21}. Even though it seems natural to comprehend the suppression by the dissipative term that, as we will see, is confirmed by the determinant of the Jacobian matrix as smaller than the unity, hence violating Liouville's theorem \cite{b7}, it is less discussed about the diffusion coefficient. We aim to fill this gap for the present model by discussing the behavior of the diffusion coefficient and its importance in suppressing unbounded diffusion. 

This paper considers a dissipative version of the standard mapping \cite{p20,b6} and discusses some aspects of the diffusive behavior \cite{p21} for an ensemble of particles moving in the phase space. We will consider a set of control parameters that leads the particles to show diffusive dynamics. However, due to the presence of dissipation, the diffusion is finite. We determine an analytical expression for the velocity behavior and an expression for the probability distribution function that gives the probability of observing a particular particle with a given action at a specific time. The diffusion coefficient is a quantity that furnishes the behavior of how easily the particles spread out over time. It quantifies the rate at which particles disperse through the phase space and gives information about the underlying mechanisms leading to the spread. We will show it is scaling invariant concerning the control parameters and time and that it turns out to be the most important result of this paper, marking our original contribution to the discussion.

The paper is organized as follows. In section \ref{sec2}, we discuss the mapping and the main essential results of its dynamics. Section \ref{sec3} discusses the solution of the diffusion equation to the universal properties and the scaling invariance observed for the diffusion coefficient. Section \ref{sec4} presents our conclusions and final remarks.

\section{The mapping and its main properties}
\label{sec2}

\begin{figure}[t]
%\vspace*{-0.1cm}
\centerline{\includegraphics[width=0.95\linewidth]{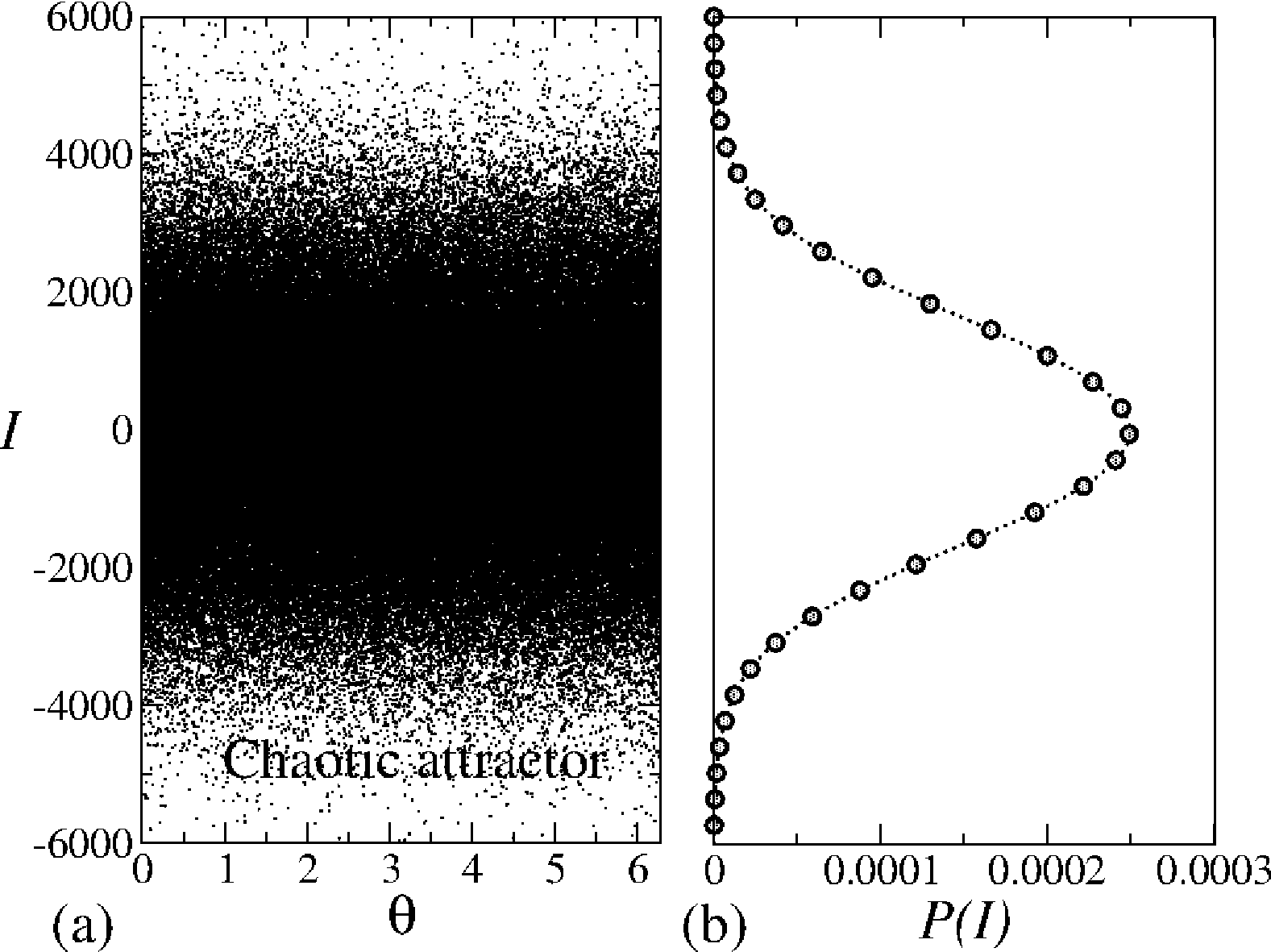}}
\caption{Plot of (a) chaotic attractor for the mapping (\ref{eq1}) considering $\epsilon=100$ and $\gamma=10^{-3}$ while (b) shows it corresponding probability distribution function giving evidence of a Gaussian distribution.}
\label{Fig0}
\end{figure}

In this section, we discuss the mapping and the main results coming from the control parameters. The Chirikov-Taylor mapping \cite{p20}, also called standard mapping \cite{b6}, is written as
\begin{equation}
T:\left\{\begin{array}{ll}
I_{n+1}=(1-\gamma)I_n+\epsilon \sin(\theta_n)\\
\theta_{n+1}=[\theta_n+I_{n+1}]~~{\rm mod (2\pi)}\\
\end{array}
\right.,
\label{eq1}
\end{equation}
where $\epsilon$ is a control parameter controlling the intensity of the nonlinearity, $\gamma\in[0,1]$ denotes the dissipation. The case of $\gamma=0$ corresponds to the conservative case, while for any $\gamma\ne 0$, the system is dissipative. The variable $I$ is the action, and $\theta$ is the angle. The operator $T$ relates a pair of variables from the time $n$ with the new pair at $(n+1)$. Some important transitions \cite{p21} for the mappings are for $\gamma=0$ and: (i) $\epsilon=0$, the system is integrable. It means the phase space is filled with periodic or quasi-periodic orbits, and no exponential divergence of different initial conditions is observed. (ii) For $\epsilon\ne 0$, the phase space is mixed, and it contains periodic orbits generally centered at the islands, chaotic sea characterized by positive Lyapunov exponents and invariant spanning curves (also called invariant tori) limiting the unbounded chaotic diffusion. (iii) For $\epsilon=\epsilon_c=0.9716\ldots$, all invariant spanning curves are destroyed \cite{b6}, and, depending on the initial conditions, the chaotic orbits are allowed to diffuse unbounded on the action axis. (iv) The case $\gamma\ne0$ brings an interesting discussion which becomes stronger with the Jacobian matrix, which is written as
$$
J=\left(\begin{array}{ll}
{{\partial I_{n+1}}\over{\partial I_n}} & {{\partial I_{n+1}}\over{\partial \theta_n}}  \\
{{\partial \theta_{n+1}}\over{\partial I_n}} & {{\partial \theta_{n+1}}\over{\partial \theta_n}}\\
\end{array}
\right),
$$
where ${{\partial I_{n+1}}\over{\partial I_n}}=(1-\gamma$), ${{\partial I_{n+1}}\over{\partial \theta_n}}=\epsilon\cos(\theta_n)$, ${{\partial \theta_{n+1}}\over{\partial I_n}}=(1-\gamma)$ and ${{\partial \theta_{n+1}}\over{\partial \theta_n}}=1+\epsilon\cos(\theta_n)$. The determinant of the Jacobian matrix is $\det J=(1-\gamma)$. For any $\gamma\ne 0$, the determinant of the Jacobian matrix is smaller than the unity, violating Liouville's theorem \cite{b7} and leading to the attractor's creation in the phase space. Figure \ref{Fig0}(a) shows a plot of the chaotic attractor observed in the phase space for the control parameters $\epsilon=100$ and $\gamma=10^{-3}$.
The probability distribution function of the points shown in Fig. \ref{Fig0}(a) is plotted in Fig. \ref{Fig0}(b) and resembles a Gaussian distribution.

\begin{figure}[t]
%\vspace*{-0.1cm}
\centerline{(a)\includegraphics[width=0.8\linewidth]{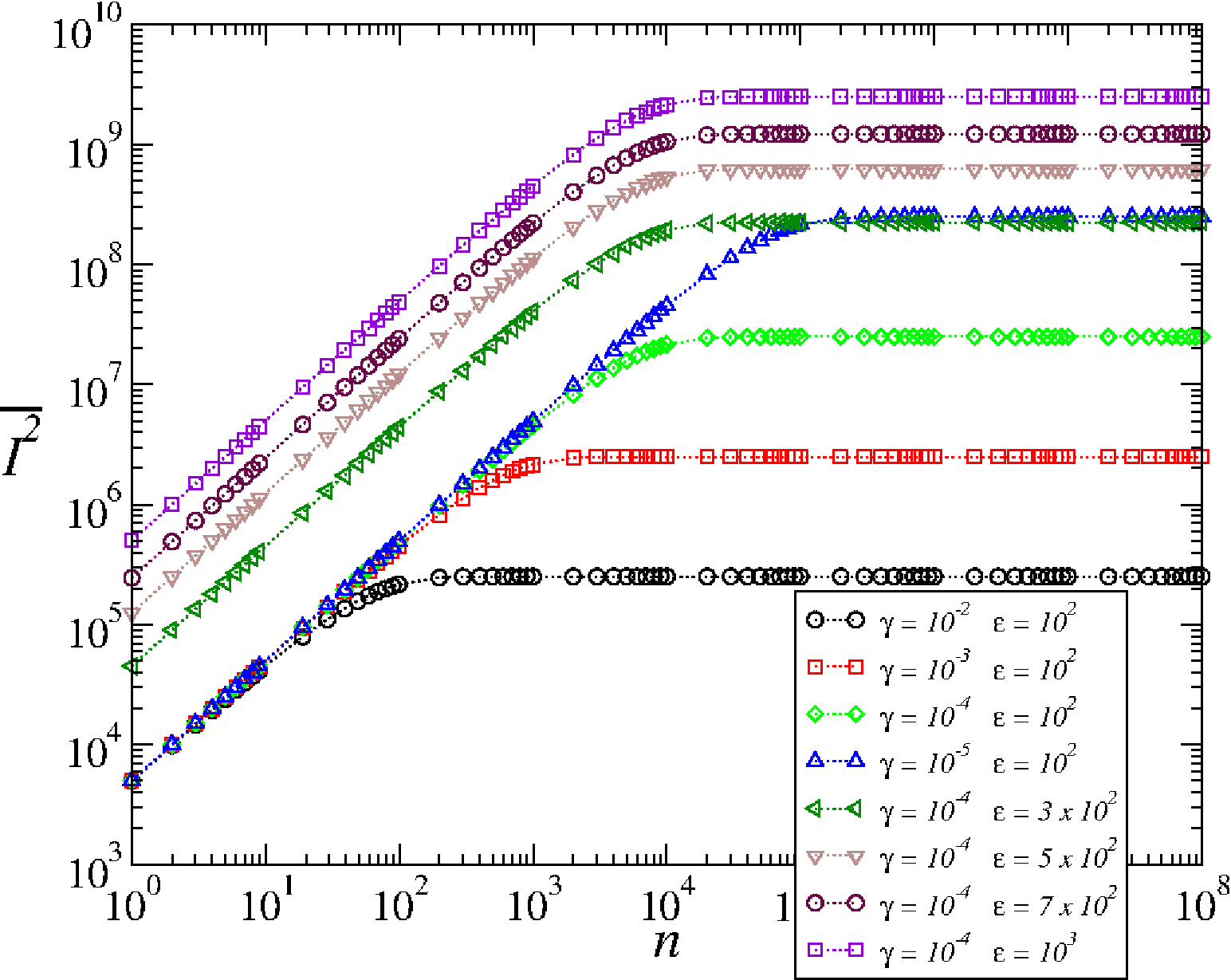}}
\centerline{(b)\includegraphics[width=0.8\linewidth]{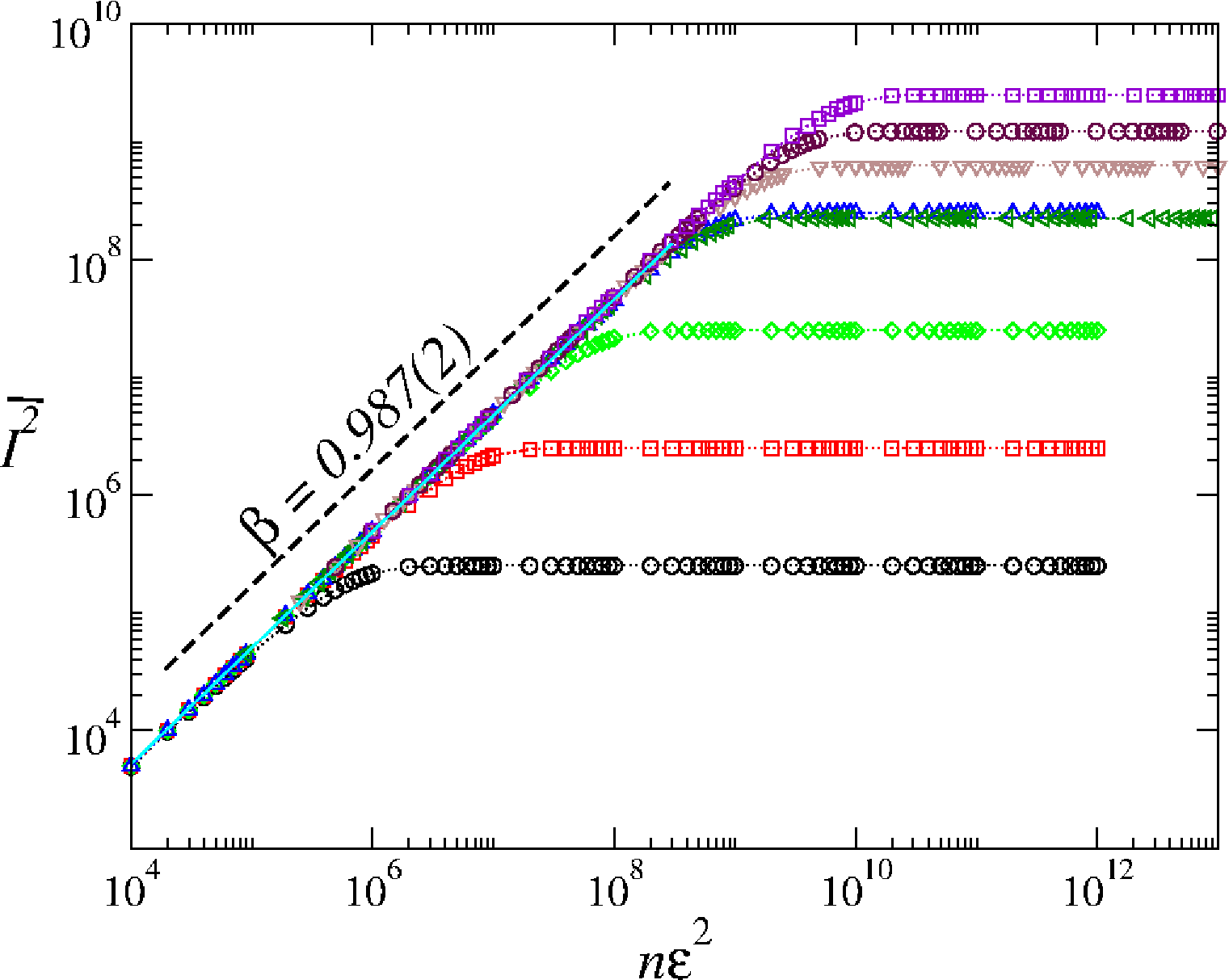}}
\caption{Plot of: (a) curves of $\overline{I^2}~vs.~n$ for different control parameters. (b) same of (a) after the transformation $n\rightarrow n\epsilon^2$ coalescing all curves to start growing together.}
\label{Fig1}
\end{figure}

Our goal is to describe how the particles diffuse in the chaotic attractor with particular attention to the diffusion coefficient and how it influences the diffusive behavior of the particles along the phase space. As we see from Fig. \ref{Fig0}(a), the distribution of points along the phase space is symmetric concerning the origin of the coordinate system. The average action $\overline{I}$ is not a good variable. Instead of it, we analyze the behavior of the squared average action. Figure \ref{Fig1}(a) shows a plot of $\overline{I^2}~vs.~n$ for different control parameters, as labeled in the figure. 

The behavior shown in Fig. \ref{Fig1} exhibits an interesting property that is commonly known as scaling invariance \cite{p21}. The typical scaling is described by the observable $I_{rms}=\sqrt{\overline{I^2}}$ and has the following properties: (i) for an initial action $I_0\approx 0$, the curves of $I_{rms}$ grows as a power law of the type $I_{rms}\propto (n\epsilon^2)^{\beta}$ for $n\ll n_x$ where $\beta=0.5$ is a critical exponent denoted as accelerating exponent and $n_x$ is a crossover time marking the changeover from the regime of growth to the saturation. (ii) for large enough time, typically $n\gg n_x$ we observe $I_{sat}\propto \epsilon^{\alpha_1}\gamma^{\alpha_2}$ where both $\alpha_i$, $i=1,2$ are critical exponents. As shown in the literature \cite{p21}, their numerical values are $\alpha_1=1$ and $\alpha_2=-1/2$. Finally, the crossover iteration number that marks the changeover from the regime of growth to the saturation is described as $n_x\propto \epsilon^{z_1}\gamma^{z_2}$ which are called dynamical critical exponents and are known as $z_1=0$ and $z_2=-1$. 

By using a homogeneous and generalized function \cite{b8} of the type $I_{rms}(n\epsilon^2,\epsilon,\gamma)=\ell I_{rms}(\ell^a n\epsilon^2,\ell^b\epsilon,\ell^c\gamma)$ where $\ell$ is a scaling factor, $a$, $b$ and $c$ are scaling exponent, a straightforward calculation gives the following scaling laws $z_1={{\alpha_1}\over{\beta}}-2$ and $z_2={{\alpha_2}\over{\beta}}$. A corresponding transformation $I_{rms}\rightarrow I_{rms}/(\epsilon^{\alpha1}\gamma^{\alpha2})$ and $n\rightarrow n/(\epsilon^{z_1}\gamma^{z_2})$ overlap all curves of $I_{rms}~vs.~n$ onto a single and universal plot. This result corroborates a scaling invariance of the diffusion along the phase space. Let us discuss how this diffusion can be described using a diffusion equation and the influence of the diffusion coefficient on the universal scaling behavior.

\section{Diffusion and scaling invariance}
\label{sec3}

Since the dynamics along the chaotic attractor resemble the dynamics of stochastic particles, we can assume that the spread of particles can be described using the diffusion equation. As quoted by the first Fick's law, a current of particles denotes a certain number of particles crossing a line in an interval of time and defines the current of particles, $J$. The current flows from a higher concentration region to a lower concentration region. It is proportional to the gradient of concentration that can be written as ${{\partial P}\over{\partial I}}$ where $P=P(I,n)$ denotes the probability of observing a particle with an action $I$ at an instant of time $n$. This proportionality can be turned into an equation by a constant $D$ called the diffusion coefficient, leading to the following equation $J=-D{{\partial P}\over{\partial I}}$ where the minus stands for the fact the current runs opposite to the potential gradient. Since the particles are still all in the dynamics and there is no loss of particles, it is immediate to use the fact there is a conservation of particles; hence ${{\partial P}\over{\partial n}}+{{\partial J}\over{\partial I}}=0$. It then leads to
\begin{equation}
{{\partial P}\over{\partial n}}={{\partial D}\over{\partial I}}{{\partial P}\over{\partial I}}+D{{\partial ^2D}\over{\partial I^2}}.
\label{eq2}
\end{equation}

The first term on the right-hand side of Eq. (\ref{eq2}) brings the dependence of $D$ concerning $I$, which is the information we do not have. As we see below, $D$ depends on the control parameters and $n$. However, its dependence is slow and smooth compared to the time variation from the state $n$ to $(n+1)$. Therefore, to solve the diffusion equation, we assume it is a constant and then replace its expression in the final result. With this assumption, the diffusion equation reads the standard view as 
\begin{equation}
{{\partial P}\over{\partial n}}=D{{\partial ^2D}\over{\partial I^2}}.
\label{eq3}
\end{equation}

Before defining the boundary and initial conditions to be used in the solution of the diffusion equation and, hence, finding a unique solution for it, let us first comment on its importance for the discussion of the observables of the problem. In statistical mechanics problems \cite{b9,b10}, the standard way to describe a given state of a system and its dynamics is by knowledge of the partition function \cite{b11}. It connects microscopical configurations of a system with properties that can be measured at a macroscopical thermodynamical view. It furnishes all possible state configurations that a system might have and, from the state equations, gives the entire description of the involved physical properties. Knowledge of its analytical expression is a desire of many scientists and researchers. In the present investigation, the probability distribution function $P(I,n)$ has the same meaning as the partition function. The analytical expression of $P(I,n)$ allows us to determine all the system's main observables and predicts their time behavior, including possible critical dynamics.

To solve the diffusion equation (\ref{eq3}) we impose that: (i) $\lim_{I\rightarrow\pm\infty}P(I,n)=0$; (ii) $P(I,0)=\delta(I-I_0)$. Condition (i) signifies all particles can not diffuse unbounded, as predicted by the determinant of the Jacobian matrix and seen from Fig. \ref{Fig0}(a), while (ii) guarantees that all particles are started together at $I=I_0$. The common technique to solve this equation is by use of a Fourier transform, as well discussed in textbooks \cite{b12,b13} that leads to the following solution
\begin{equation}
P(I,n)={{1}\over{\sqrt{4\pi Dn}}}e^{-{{(I-I_0)^2}\over{4Dn}}}.
\label{eq4}
\end{equation}
It then gives the probability of observing a particular particle at a specific action $I$ in the time $n$. All momenta of the distribution can be obtained directly from $P(I,n)$.

Let us now properly discuss the behavior of the diffusion coefficient $D$. It is defined as the mean squared displacement and is written as $D=(\overline{I^2}_{n+1}-\overline{I^2}_n)/2$. To obtain an expression for $D$, we have to know the behavior of $\overline{I^2}_{n+1}$ and $\overline{I^2}_n$, that can be obtained from the first equation of the mapping (\ref{eq1}). Taking the square of first equation we end up with $I^2_{n+1}=(1-\gamma)^2I^2_n+\epsilon^2\sin^2(\theta_n)+2(1-\gamma)\epsilon I_n\sin(\theta_n)$ and assuming statistical independence of $I$ and $\theta$ at the chaotic attractor, an ensemble average gives
\begin{equation}
\overline{I^2}_{n+1}=(1-\gamma)^2\overline{I^2}_n+{{\epsilon^2}\over{4}}.
\label{eq5}
\end{equation}
Substituting this result in the equation of $D$ we obtain $D={{\gamma(\gamma-2)\overline{I^2}_n}\over{2}}+{{\epsilon^2}\over{4}}$ that makes us to obtain the behavior of $\overline{I^2}_n$. It can be obtained from the following assumption
\begin{equation}
\overline{I^2}_{n+1}-\overline{I^2}_n={{\overline{I^2}_{n+1}-\overline{I^2}_n}\over{(n+1)-n}}\approx {{d\overline{I^2}}\over{dn}}.
\label{eq6}
\end{equation}
An immediate integration for Eq. (\ref{eq6}) gives
\begin{equation}
\overline{I^2}(n)=I^2_0e^{-\gamma(2-\gamma)n}+{{\epsilon^2}\over{2\gamma(\gamma-2)}}\left[e^{-\gamma(2-\gamma)n}-1\right].
\label{eq7}
\end{equation}
It has important limits: (i) $n=0$ gives $\overline{I^2(0)}=I^2_0$; (ii) for the case of $n\rightarrow\infty$ we obtain
\begin{equation}
\overline{I^2}_{sat}={{\epsilon^2}\over{2\gamma(2-\gamma)}}.
\label{eq8}
\end{equation}
This result confirms two critical exponents as $\alpha_1=1$ and $\alpha_2=-1/2$. From the scaling laws, we also obtain $z_1=0$ and $z_2=-1$ for the case of $\beta=1/2$.

Substituting Eq. (\ref{eq7}) in the expression of $D$ and after the simplifications we obtain
\begin{equation}
D(n)=\left[{{I^2_0\gamma(\gamma-2)}\over{2}}+{{\epsilon^2}\over{4}}\right]e^{-\gamma(2-\gamma)n},
\label{eq9}
\end{equation}
which also have important limits. For the case of $n=0$, we recover the well know result $D={{\epsilon^2}\over{4}}$ when $I_0\cong 0$. The other limit is for $n\rightarrow\infty$, leading to $D(\lim n\rightarrow\infty)=0$. It implies the particle is no longer experiencing diffusion, leading the dynamics to the stationary state.

\begin{figure}[t]
%\vspace*{-0.1cm}
\centerline{\includegraphics[width=1.0\linewidth]{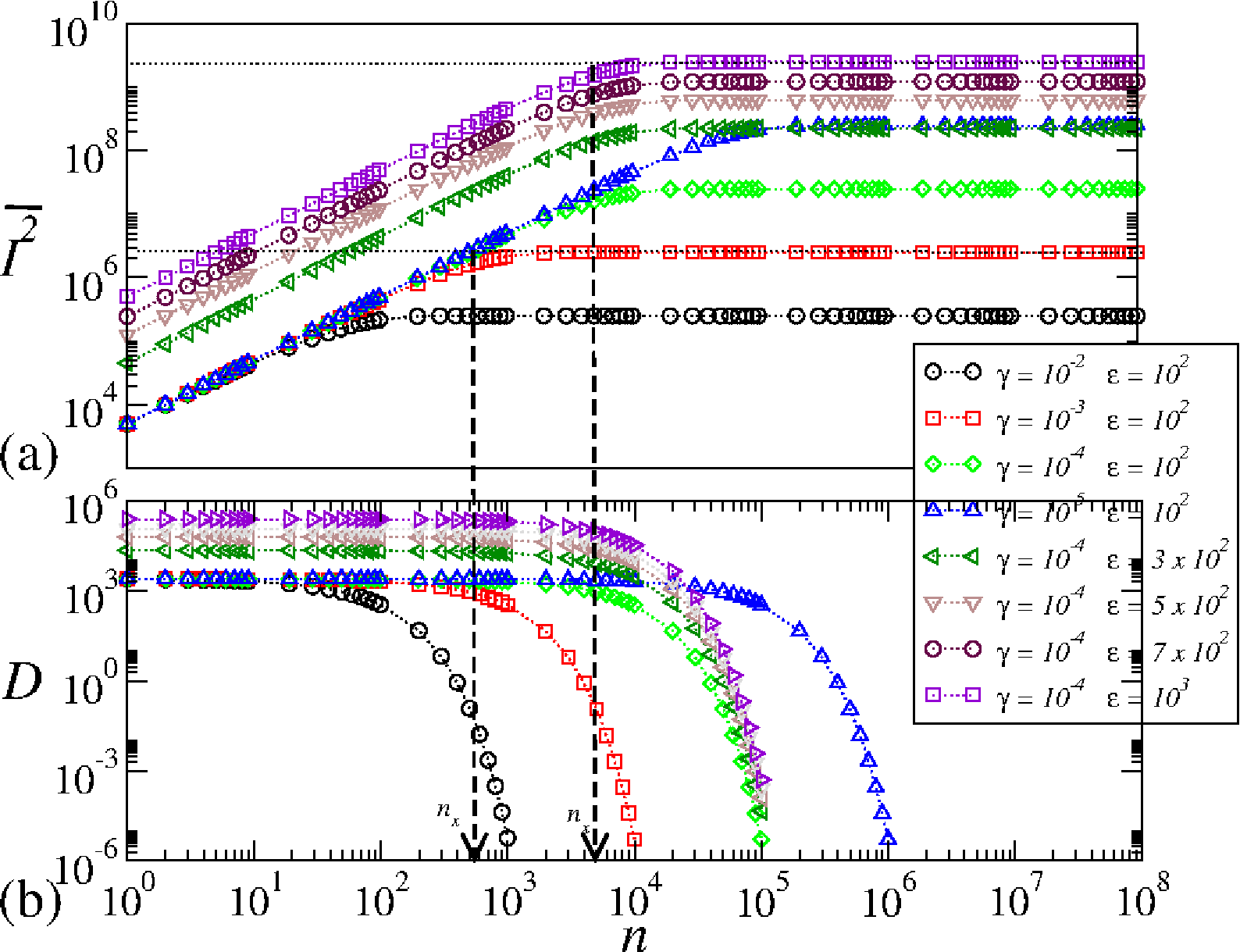}}
\caption{(Color online) Plot of: (a) curves of $\overline{I^2}~vs.~n$ for different control parameters. (b) $D~vs.~n$ for the same control parameters of (a).}
\label{Fig2}
\end{figure}

A plot of $\overline{I^2}~vs.~n$ and $D~vs.~n$ is shown in Fig. \ref{Fig2}(a-b) for different control parameters, as labeled in the figure.
The curves of $D$ remain constant for a short time, corresponding to the regime of diffusion for the action variables, leading to growth for the curves of $\overline{I^2}$. Eventually, $D$ passes from a crossover and bends towards a decay regime. The crossover of the constant plateau to the decay dynamics coincides nicely with the changeover from the growth domain of diffusion to saturation.

\begin{figure}[t]
%\vspace*{-0.1cm}
\centerline{(a)\includegraphics[width=0.8\linewidth]{Fig3a.eps}}
\centerline{(b)\includegraphics[width=0.8\linewidth]{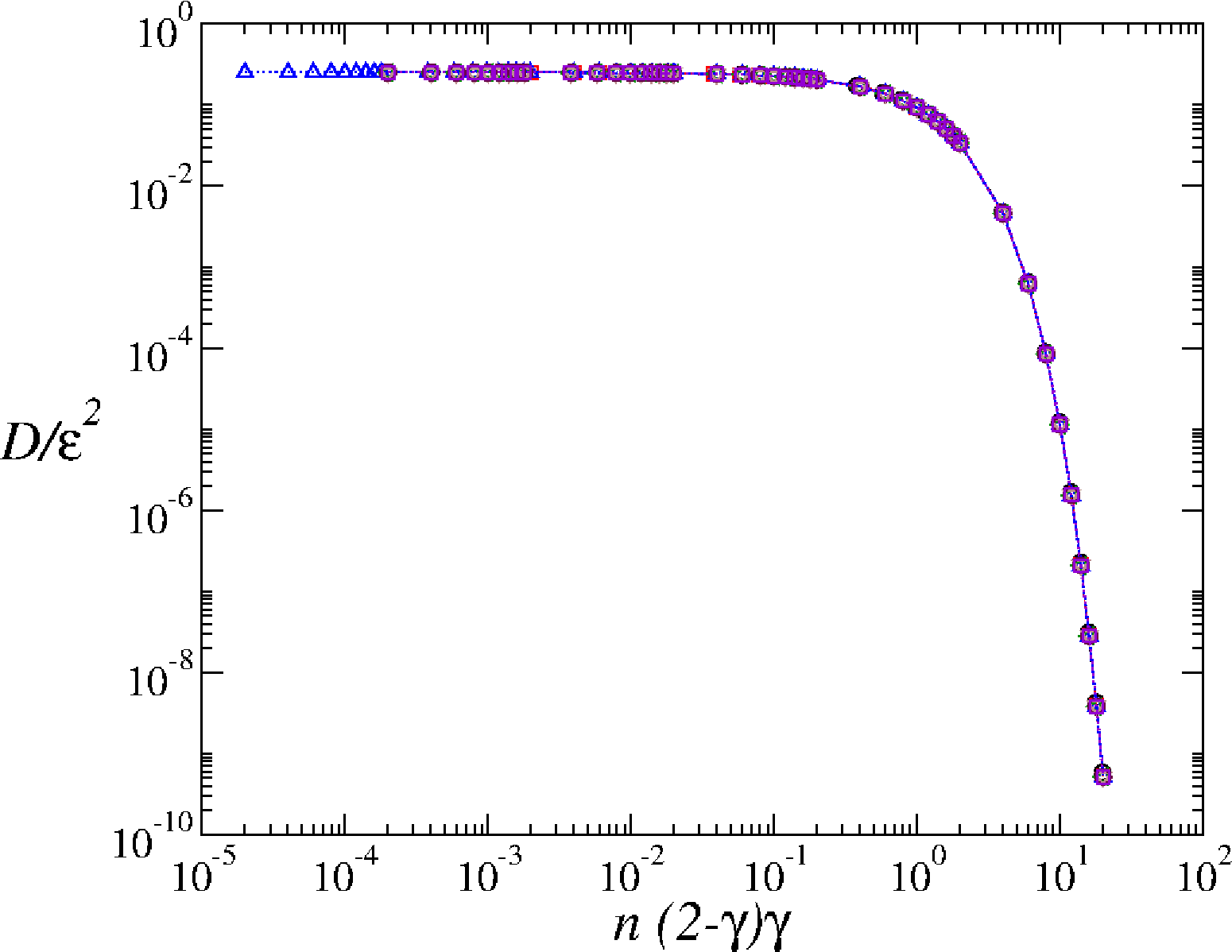}}
\caption{Plot of: (a) curves of $D~vs.~n$ for different control parameters. (b) same of (a) after the transformation $D\rightarrow D/\epsilon^2$ and $n\rightarrow n\gamma(2-\gamma)$, overlapping all curves shown in (a) onto a single and universal plot.}
\label{Fig3}
\end{figure}

If we Taylor expand Eq. (\ref{eq9}) considering the case of $I_0\cong 0$, we obtain
\begin{equation}
D(n)={{\epsilon^2}\over{4}}\left[1+\gamma(-2+\gamma)n+{{1}\over{2}}\gamma^2(-2+\gamma)^2n^2+\ldots\right].
\label{eq10}
\end{equation}
The first dominant term marking the crossover from a constant plateau to a regime of decay can be estimated when $1+\gamma(-2+\gamma)n_x\cong 0$, leading to
\begin{equation}
n_x={{1}\over{2-\gamma}}\gamma^{-1}. 
\label{eq11}
\end{equation}
The crossover for the plateau for the regime of decay is described by the same critical exponent $z_2=-1$. We can also note that the scaling variables are then: (i) $D\rightarrow D/\epsilon^2$ and (ii) $n\rightarrow n \gamma(2-\gamma)$. Figure \ref{Fig3}(a) shows a plot of $D~vs.~n$ for different control parameters, while Fig. \ref{Fig3}(b) shows the overlap of the curves shown in (a) after the scaling transformations.

It is interesting to note that each control parameter is essential for the dynamics. The control parameter $\epsilon$ affects the diffusion length responsible for the particles' elementary excitation, leading them to diffuse. However, the parameter $\gamma$ marks how long the diffusion along the phase space is, hence limiting the diffusion and setting a bound for it. Another important point is that the assumption considered for the solution of the diffusion equation is true in the sense the diffusion coefficient $D$ variation is slow from the instant $n$ to $(n+1)$. At this limit, it can be considered constant for the solution of the diffusion equation and does not affect the characterization of the main observables of the system.

\section{Summary and conclusions}
\label{sec4}
We discussed the behavior of the diffusion coefficient for a transition from bounded to unbounded diffusion in a dissipative standard mapping. The dynamics for a dissipative case violate de Liouville's theorem and create attractors in the phase space since the determinant of the Jacobian matrix is smaller than the unity. Since the attractors are far away from the infinity, unbounded growth of the action is not observed. The probability density to observe a particle with a specific action at an instant of time was obtained by the solution of the diffusion equation. We assumed the diffusion coefficient varied slowly in time, and at that point, it can be considered constant from the instant $n$ to $(n+1)$. From the definition of the diffusion coefficient, we obtained an analytical expression for it using the mean squared displacement. We proved that scaling is invariant to the control parameter. The diffusion coefficient remains constant for a short time and starts with an initial action close to zero, leading the particles to diffuse. Suddenly, it changes from a plateau to a regime of decay. The changeover from the plateau to the decay scales with $n_x={{\gamma^{-1}}\over{2-\gamma}}$. The exponent $-1$ for $n_x$ is the same as observed for the crossover in the dissipative time dependent billiard \cite{p22,p23}, hence giving arguments the transition from bounded to the unbounded diffusion in the dissipative standard mapping can be fitted in the same universality class of the dissipative time dependent billiard. Moreover, unbounded diffusion for the action is suppressed when the coefficient diffusion goes to zero.

\section*{Acknowledgements}
CMK thanks to CAPES for support E.D.L. acknowledges support from Brazilian agencies CNPq (No. 301318/2019-0, 304398/2023-3) and FAPESP (No. 2019/14038-6 and No. 2021/09519-5).

\end{document}